\documentclass[copyright,creativecommons]{eptcs}
\providecommand{\event}{}
\providecommand{\volume}{2}
\providecommand{\anno}{2009}
\providecommand{\firstpage}{115}
\usepackage{breakurl}  
\usepackage{semantic}
\usepackage{wasysym}
\usepackage{array}

\title{Fairness as a QoS Measure for Web Services}
\author{Stefano Bistarelli
\institute{Dipartimento di Scienze, Universit\`a ``G. d'Annunzio''
\\Chieti-Pescara, Italy} \institute{Dipartimento di Matematica
Informatica, \\ Universit\`a di Perugia, Italy} \institute{ Istituto
di Informatica e Telematica (CNR),\\ Pisa, Italy}
\email{bista@dipmat.unipg.it} \and Paola Campli
\institute{Dipartimento di Scienze, Universit\`a ``G. d'Annunzio''
\\Chieti-Pescara, Italy} \email{campli@sci.unich.it} }

\def\titlerunning{Fairness as a QoS Measure for Web Services}
\def\authorrunning{S.~Bistarelli \& P.~Campli}
\begin{document}
\maketitle

\begin{abstract}
Service Oriented Architectures (SOAs) are component-based
architectures, characterized by reusability, modularization and
composition, usually offered by HTTP (web services) and often
equipped with a Quality of Services (QoS) measure. In order to
guarantee the fairness property to each client requesting a service,
we propose  a fair version of the (Soft) Concurrent Constraint
language to deal with the negotiation phases of the Service Level
Agreement (SLA) protocol.
\end{abstract}

\section{Introduction}
Service Oriented Architecture (SOAs) are the current technology for
business to business interactions.
 SOAs are component-based architectures, characterized by  reusability,
 modularization and composition, usually  offered via HTTP
 (web services) and often equipped with a \textit{Quality of Service}
  (QoS) measure. \\
For example if the required service is routing a possible measure of
quality could be bandwidth, cost or delay. However these metrics do
not reflect the subjective criteria of individual users or groups.
In this article, we consider \emph{fairness} as a new parameter to
be adopted in a QoS measure and we propose to use Soft Constraint to
represent the QoS scores as suggested in~\cite{bista-sant}.\\
When a user requests a service, he evaluates it with a QoS measure;
we think that fairness is an important parameter to consider, as it
directly affects the QoS value; therefore the service provider aims
to guarantee a fair service in order to obtain a high QoS score.\\
We use (Soft) Concurrent Constraint (SCC) programming to model web
services requests. Based on this formalism we introduce \emph{Fair
Concurrent Constraint Programming}, that uses an extension of the
parallel operator which enables the provider to guarantee the
fairness property to each client requesting a service. We consider
soft constraints in order to define fairness as the level of
preference associated to each constraint.

This paper is organized as follow: In section 2 we give a background
of the Service Oriented Architecture; in section 3 we explain how to
apply the soft concurrent constraint  framework to the QoS
negotiation; in section 4 we define the various notions of fairness;
in section 5 we propose a fair version of the concurrent constraint
(cc) language. Finally, in section 6 we model the SLA negotiation
with the soft concurrent constraint language.

\section{Service Oriented Architecture }

A Service Oriented Architecture can be defined as a group of
services, which communicate with each other~\cite{soa}. The process
of communication involves either simple data passing or it could
involve two or more services coordinating some activities. Some
means of connecting services to each other are needed. Basic
services, their descriptions, and basic operations (publication,
discovery, selection, and binding) that produce or utilize such
descriptions constitute the SOA foundation. The higher layers in the
SOA pyramid provide additional support required for service
composition and service management. The service composition layer
encompasses necessary roles and functionality for the consolidation
of multiple services into a single composite service. The resulting
composite services may be used by service aggregators as components
(basic services) in further service compositions or may be utilized
as applications/solutions by service clients. The main
characteristic of a SOAs are four:
\begin{itemize}
\item Coordination: Controls the execution of component services, and manages dataflow among them and
to the output of the component service (by specifying workflow
processes and using a workflow engine for runtime control of service
execution).
\item Monitoring: Subscribes to events or information produced by the component services, and publishes
higher-level composite events (by filtering, summarizing, and
correlating component events).
\item Conformance: Ensures the integrity of the composite service by matching its parameter types with
those of its components, imposes constraints on the component
services (to ensure enforcement of business rules), and performing
data fusion activities.
\item Quality of Service (QoS) composition: Leverages, aggregates, and bundles the component's QoS to
derive the composite QoS, including the composite service's overall
cost, performance, security, authentication, privacy,
(transactional) integrity, reliability, scalability, and
availability.
\end{itemize}
Usually QoS metrics do not reflect the subjective criteria of
individual users or groups. Our goal is to extend such QoS measure
by adding Fairness as a new parameter in order to consider a
\emph{subjective perception} of the quality of the service. A
subjective quality of the service represent the ``ethical acceptance
of the service" and it cannot be measured by the service provider
just looking at one of the user, since all the users need to be
considered together.

\section{Managing QoS Measures with Soft Concurrent Constraint}
The language we will use for web service interaction, composition
and contracting is the Soft Concurrent Constraint~\cite{bistabook}.
Constraints has been used successfully to represent service
interactions, contract requirements, and general services desiderata~\cite{lazovik}.


The aim of the paper is to apply  \emph{Quality of Service} (QoS)
measures for \emph{Service Oriented Architectures} considering
Fairness as a new subjective attribute for QoS and for representing
contracts and \emph{Service Level Agreements} (SLAs)~\cite{sla}  in
terms of constraint based languages. The notions of contract and
SLAs are very important in SOC (service oriented computing) since
they allow to describe the mutual interaction between communicating
parties and to express properties related to the quality of service
such as cost, performance, reliability and availability.

The use of a soft constraint framework permits to manage SOAs in a
declarative fashion by considering together both the
requirements/interfaces of each service and their QoS estimation~\cite{bista-sant}. C-semirings can represent several QoS attributes,
while soft constraints represent the specification of each service
to integrate: they link these measures to the resources spent in
providing it, for instance, ``the reliability is equal to 80\% plus
5\% for each other processor used to execute the service''. This
statement can be easily represented with a soft constraint where the
number of processors corresponds to the $x$ variable, and the
preference (i.e.~reliability) level is given by the $5x + 80$
polynomial. In the same way we can consider and deal with the
fairness property within QoS.

A Soft Constraint-based formal language permit us to model and check
QoS. In order to understand how to use the soft concurrent
constraint framework we explain the key concepts related to it in
the following sections.

\subsection{Concurrent Constraint Programming}
The concurrent constraint (cc) programming paradigm~\cite{cc}
 concerns the behavior of a set of concurrent
agents with a shared store, which is a conjunction of constraints (
relations among a specified set of variables). Each computation step
possibly adds new constraints to the store. Thus information is
monotonically added until all agents have evolved. The final store
is a refinement of the initial one and it is the result of the
computation. The concurrent agents communicate with the shared
store, by either checking if it entails a given constraint ({\em
ask} operation) or adding a new constraint to it ({\em tell}
operation). \\ The syntax of a cc program is show in
Table~\ref{tab:cc}: $P$ is the class of programs, $F$ is the class
of sequences of procedure declarations (or clauses), $A$ is the
class of agents, $c$ ranges over constraints, and $x$ is a tuple of
variables. The $+$ combinator express nondeterminism. We also assume
that, in $p(x) :: A, vars(A) \subseteq x$, where $vars(A)$ is the
set of all variables occurring free in agent $A$. In a program $P =
F.A$, $A$ is the initial agent,
to be executed in the context of the set of declarations $F$.\\
\begin{table}[hbt]
\centering{\caption{cc syntax.} \label{tab:cc} \hrule \vskip2pt
\begin{eqnarray*}
P & ::= & F.A\\
F & ::= & p(x)::A \mid F.F\\
A & ::= & success \mid fail \mid tell(c) \rightarrow A \mid E \mid A
\| A \mid \exists_x A \mid p(x)\\
E & ::= & ask(c) \rightarrow A \mid E+E
\end{eqnarray*} \vskip2pt
\hrule
}
\end{table}

The intuitive behavior of the agents is: agent ``$success$''
succeeds in one step; agent ``$fail$'' fails in one step; agent
``$ask(c) \rightarrow A$'' checks whether constraint $c$
  is entailed by the current store and then, if so, behaves like agent
  $A$. If $c$ is inconsistent with the current store, it fails, and
  otherwise it suspends, until $c$ is either entailed by the current
  store or is inconsistent with it; agent ``$ask(c_1) \rightarrow A_1 + ask(c_2) \rightarrow A_2$''
  may behave either like $A_1$ or like $A_2$ if both $c_1$ and $c_2$ are
  entailed by the current store, it behaves like $A_i$ if $c_i$ only
  is entailed, it suspends if both $c_1$ and $c_2$ are consistent with
  but not entailed by the current store, and it behaves like
  ``$ask(c_1) \rightarrow A_1 $'' whenever ``$ask(c_2 ) \rightarrow
  A_2$'' fails (and vice versa);
 agent ``$tell(c) \rightarrow A$'' adds constraint $c$ to the current store
  and then, if the resulting store is consistent, behaves like $A$,
  otherwise it fails;
 agent $A_1 \| A_2$ behaves like $A_1$ and $A_2$ executing
  in parallel;
agent $\exists_x A$ behaves like agent $A$, except that the
variables in $x$ are local to $A$;
 $p(x)$ is a call of procedure $p$.

Here is a brief description of the transition rules:

\begin{center}
\begin{tabular} {lr}
\hline
\\
 $ \langle tell(c) \to  A, \sigma \rangle   \to  \langle A,
\sigma \otimes c \rangle $
&  tell \\

\\
$ \frac{\sigma \vdash c, } { \langle Ask(c) \to  A, \sigma \rangle
\to \langle  A, \sigma \rangle} $
&  ask \\

\\
$ \frac  { \langle A_{1}, \sigma \rangle   \to   \langle A_{1}^{'},
\sigma^{'} \rangle } {\langle A_{1}
\parallel A_{2},\sigma \rangle   \to  \langle
A^{'}_{1}\parallel A_{2}, \sigma^{'} \rangle }  $

$ \frac  { \langle A_{1}, \sigma \rangle   \to   \langle A_{1}^{'},
\sigma^{'} \rangle } { \langle A_{2}\parallel A_{1},\sigma \rangle
\to  \langle A_{2}\parallel A^{'}_{1}, \sigma^{'} \rangle}   $

& parallelism (1)\\
\\
$ \frac{ \langle A_{1}, \sigma \rangle    \to   \langle success,
\sigma ^{'} \rangle  } { \langle A_{1}\parallel A_{2}, \sigma
\rangle\to
 \langle A_{2}, \sigma^{'} \rangle} $

$ \frac{ \langle A_{1}, \sigma \rangle    \to   \langle success,
\sigma ^{'} \rangle  } {\langle A_{2}\parallel  A_{1}, \sigma
\rangle  \to  \langle A_{2}, \sigma^{'} \rangle  } $

&  parallelism (2)\\
\\
$
 \frac{ \langle  E_{1}, \sigma \rangle   \to  \langle  A_{1},
\sigma^{'} \rangle  }{\langle  E_{1}+E_{2}, \sigma \rangle \to
\langle  A_{1}, \sigma^{'} \rangle  }  $
 $
 \frac{ \langle  E_{1}, \sigma \rangle   \to  \langle  A_{1},
\sigma^{'} \rangle  }{ \langle E_{2}+E_{1}, \sigma \rangle  \to
\langle A_{1}, \sigma^{'} \rangle } $
&  nondeterminism \\

\\
$ \frac{ \langle  A[y/x], \sigma \rangle   \to  \langle A_{1},
\sigma{'} \rangle  } { \langle  \exists_{x} A , \sigma \rangle   \to
\langle  A_{1}, \sigma^{'} \rangle  } $

&  hidden variables \\
\\
 $ \langle  p(y), \sigma \rangle   \to  \langle  A[y/x],
\sigma \rangle \quad \textrm{where} \quad p(x) :: A $

&  procedure call \\
\\
\hline
\end{tabular}
\end{center}

\subsection{Soft Constraint}
Soft constraints generalize classical constraints by allowing
several levels of satisfaction, and are able to express within the
constraints many optimization criteria, and even the combination of
some of them. The framework of soft constraints based on semirings~\cite{bistabook} has shown to be general and expressive enough to
deal with fuzziness, uncertainty, multiple criteria and also partial
orders of preference levels. \\
The features of soft constraints could be useful in representing
routing problems where an imprecise state information is given.
Moreover, since QoS is only a specific application of a more general
notion of Service Level Agreement (SLA), many applications could be
enhanced by using such a framework. As an example consider
E-commerce: here we are always looking for establishing an agreement
between a merchant and a client. Also, all auction-based
transactions need an agreement protocol.
\\
Moreover, the possibility to embed soft constraints in a concurrent
language environment is very useful to model agents interaction over
the web. In fact, in this way not only we model the way constraints
are solved, but we also have the possibility to describe the
inter-agent protocols that are used to reach a common agreement.

For these reasons, in this paper we use the soft extension of the
concurrent constraint language~\cite{cc} with the possibility to
handle soft constraints. \\In the soft concurrent constraint, tell
and ask agents are equipped with a preference (or consistency)
threshold which is used to determine their success, failure, or
suspension. The use of this thresholds yields an interesting and
useful computational model, where it is possible to reason about the
most convenient preference threshold to use to find the whole
observable semantics of the language.

\subsection{Soft Constraint Problems}

A soft constraint may be seen as a constraint where each
instantiations of its variables has an associated value from a
partially ordered set. Combining constraints will then have to take
into account such additional values, and thus the formalism has also
to provide suitable operations for combination ($\times$) and
comparison ($+$) of tuples of values and constraints. This is why
this formalization is based on the concept of c-semiring, which is
just a set plus two operations.

\paragraph{c-semirings.}
A semiring is a tuple $\langle A,+,\times,0,1 \rangle$ such that:
\begin{itemize}
\item $A$ is a set and $0, 1 \in A$;
\item $+$ is commutative, associative and $0$ is its unit element;
\item $\times$ is associative, distributes over $+$, $1$  is its
unit element and $0$ is its absorbing element.
\end{itemize}
A {\em c-semiring} is a semiring $\langle A,+,\times,0,1 \rangle$
such that: $+$ is idempotent, $1$ as its absorbing element and
$\times$ is commutative.

\paragraph{Soft Constraint Systems.}
A {\em soft constraint system} is a tuple $CS= \langle S, D, V
\rangle$ where $S$ is a c-semiring, $D$ is a finite set (the domain
of the variables) and $V$ is an ordered set of variables. Given a
semiring $S = \langle A,+,\times,0,1 \rangle$ and a constraint
system $CS= \langle S, D, V \rangle$, a {\em constraint} is a pair
$\langle def, con \rangle$ where $con \subseteq V$ and $def:
D^{|con|} \rightarrow A$. Therefore, a constraint specifies a set of
variables (the ones in $con$), and assigns to each tuple of values
of these variables an element of the semiring. Consider two
constraints $c_1=\langle def_1, con \rangle$ and $c_2=\langle def_2,
con \rangle$, with $|con|=k$. Then $c_1 \sqsubseteq_S c_2$ if for
all k-tuples $t$, $def_1(t) \leq_S def_2(t)$. The relation
$\sqsubseteq_S$ is a partial order.

A {\em soft constraint problem} is a pair $\langle C, con \rangle$
where $con \subseteq V$ and $C$ is a set of constraints: $con$ is
the set of variables of interest for constraint set $C$, which
however may concern also variables not in $con$. Note that a
classical CSP is a SCSP where the chosen c-semiring is: $S_{CSP} =
\langle \{false, true\},\vee, \wedge, false, true \rangle$. Fuzzy
CSPs  can instead be modeled in the SCSP framework by choosing the
c-semiring $S_{FCSP} = \langle [0,1], max, min, 0, 1 \rangle$.

\paragraph{Combining and projecting soft constraints.} Given the two
constraints $c_1 = \langle def_1,con_1 \rangle$ and $c_2 = \langle
def_2,con_2 \rangle$, their {\em combination} $c_1 \otimes c_2$ is
the constraint $\langle def,con \rangle$ defined by $con = con_1
\cup con_2$ and $def(t) = def_1(t \downarrow^{con}_{con_1}) \times
def_2(t \downarrow^{con}_{con_2})$, where $t \downarrow^X_Y$ denotes
the tuple of values over the variables in $Y$, obtained by
projecting tuple $t$ from $X$ to $Y$. In words, combining two
constraints means building a new constraint involving all the
variables of the original ones, and which associates to each tuple
of domain values for such variables a semiring element which is
obtained by multiplying the elements associated by the original
constraints to the appropriate subtuples.

Given a constraint $c = \langle def,con \rangle$ and a subset $I$ of
$V$, the {\em projection} of $c$ over $I$, written $c \Downarrow_I$
is the constraint $\langle def', con' \rangle$ where $con' = con
\cap I$ and $def'(t') = \sum_{t / t \downarrow^{con}_{I \cap con} =
t'} def(t)$.
Informally, projecting means eliminating some variables. This is
done by associating to each tuple over the remaining variables a
semiring element which is the sum of the elements associated by the
original constraint to all the extensions of this tuple over the
eliminated variables. In short, combination is performed via the
multiplicative operation of the semiring, and projection via the
additive operation.

\paragraph{Solutions.} The {\em solution} of an SCSP problem
${\mathsf P} = \langle C,con \rangle$ is the constraint
$Sol({\mathsf P})=(\bigotimes C) \Downarrow_{con}$.
That is, we combine all constraints, and then project over the
variables in $con$. In this way we get the constraint over $con$
which is ``induced'' by the entire SCSP.

Sometimes it may be useful to find the semiring value of the optimal
solutions. This is called the {\em best level of consistency} of an
SCSP problem $P$ and it is defined by $blevel(P) = Sol(P)
\Downarrow_{\emptyset}$. We also say that: $P$ is
$\alpha$-consistent if $blevel(P) = \alpha$; $P$ is consistent iff
there exists $\alpha >_S 0$ such that $P$ is $\alpha$-consistent;
$P$ is inconsistent if it is not consistent.

\subsection{Soft Concurrent Constraint}
We briefly explain the main differences between CC and SCC~\cite{scc}: \\
In the soft framework there are no notions of consistency and
inconsistency. Instead, a notion of {\em $\alpha$-consistency} is
introduced. This means that the syntax and semantics of the tell and
ask agents have to be enriched with a threshold to specify when
ask/tell agents have to fail, succeed or suspend.

Moreover, the notion of a thresholds $\alpha$ could be given
pointwise w.r.t. each variable assignment (this is the notion of the
following $\rightarrow_{\phi}$) or w.r.t. an overall consistency
level of the constraint store (leading to the notion of
$\rightarrow^a$).
\\
Given a soft constraint system $\langle S,D,V\rangle$, the
corresponding structure $C$, and any constraint $\phi \in C$, the
syntax of {\em scc} agents is given in Table~\ref{tab:scc}.

\begin{table}[hbt]
\caption{scc syntax.} \label{tab:scc} \hrule \vskip2pt
\begin{eqnarray*}
P & ::= & F.A\\
F & ::= & p(X)::A \mid F.F\\
A & ::= & stop \mid tell(c) \rightarrow_{\phi} A \mid tell(c)
\rightarrow^a A \mid E \mid A \| A
\mid \exists X.A \mid p(X)\\
E & ::= & ask(c) \rightarrow_{\phi} A \mid ask(c) \rightarrow^a A \mid
E+E
\end{eqnarray*} \vskip2pt
\hrule
\end{table}

The Operational Semantic for scc programs is given in the SOS style:
\begin{center}
\begin{tabular} {lr}
\hline
\\
 $\frac{(\sigma \otimes c)\Downarrow_{\emptyset} \not< a}{ \langle tell(c) \to^{a}  A, \sigma \rangle   \to  \langle A,
\sigma \otimes c \rangle} $
& valued tell \\
\\
 $\frac{\sigma \otimes c \not\sqsubset \phi}{ \langle tell(c) \to_{\phi}  A, \sigma \rangle   \to  \langle A,
\sigma \otimes c \rangle} $
& tell \\

\\
$ \frac{\sigma \vdash c, \sigma\Downarrow_{\emptyset} \not< a} {
\langle ask(c) \to^{a}  A, \sigma \rangle \to \langle  A, \sigma
\rangle} $
&  valued ask \\

\\
$ \frac{\sigma \vdash c, \sigma \not\sqsubset \phi} { \langle ask(c)
\to_{\phi}  A, \sigma \rangle \to \langle  A, \sigma \rangle} $
&  ask \\

\\
$ \frac  { \langle A_{1}, \sigma \rangle   \to   \langle A_{1}^{'},
\sigma^{'} \rangle } {\langle A_{1}
\parallel A_{2},\sigma \rangle   \to  \langle
A^{'}_{1}\parallel A_{2}, \sigma^{'} \rangle }  $

$ \frac  { \langle A_{1}, \sigma \rangle   \to   \langle A_{1}^{'},
\sigma^{'} \rangle } { \langle A_{2}\parallel A_{1},\sigma \rangle
\to  \langle A_{2}\parallel A^{'}_{1}, \sigma^{'} \rangle}   $

& parallelism (1)\\
\\
$ \frac{ \langle A_{1}, \sigma \rangle    \to   \langle success,
\sigma ^{'} \rangle  } { \langle A_{1}\parallel A_{2}, \sigma
\rangle\to
 \langle A_{2}, \sigma^{'} \rangle} $

$ \frac{ \langle A_{1}, \sigma \rangle    \to   \langle success,
\sigma ^{'} \rangle  } {\langle A_{2}\parallel  A_{1}, \sigma
\rangle  \to  \langle A_{2}, \sigma^{'} \rangle  } $

&  parallelism (2)\\
\\
$
 \frac{ \langle  E_{1}, \sigma \rangle   \to  \langle  A_{1},
\sigma^{'} \rangle  }{\langle  E_{1}+E_{2}, \sigma \rangle \to
\langle  A_{1}, \sigma^{'} \rangle  }  $
 $
 \frac{ \langle  E_{1}, \sigma \rangle   \to  \langle  A_{1},
\sigma^{'} \rangle  }{ \langle E_{2}+E_{1}, \sigma \rangle  \to
\langle A_{1}, \sigma^{'} \rangle } $
&  nondeterminism \\
\\
$ \frac{ \langle  A[y/x], \sigma \rangle   \to  \langle A_{1},
\sigma{'} \rangle  } { \langle  \exists_{x} A , \sigma \rangle   \to
\langle  A_{1}, \sigma^{'} \rangle  } $

&  hidden variables \\
\\
 $ \langle  p(y), \sigma \rangle   \to  \langle  A[y/x],
\sigma \rangle \quad \textrm{when} \quad p(x) :: A $

&  procedure call \\
\\
\hline
\end{tabular}
\end{center}

Here is a brief description of the rules:
\begin{description}
  \item[Stop] The stop agent succeeds in one step by transforming
  itself into terminal configuration {\em success}.
  \item[Valued-tell] The valued-tell rule checks for the $\alpha$-consistency
  of the SCSP defined by the store $\sigma \cup c$. The rule can
  be applied only if the store $\sigma \cup c$ is $b$-consistent
  with $b \not\leq a$. In this case the agent evolves to the new
  agent $A$ over the store $\sigma \otimes c$.
  Note that different
  choices of the {\em cut level} $a$ will lead to different
  computations.
  \item[Tell] The tell action is a finer check of the
  store. In this case, a pointwise comparison between the store
  $\sigma \otimes c$ and the constraint $\phi$
  is performed. The idea is to perform an overall
  check of the store and to continue the computation only if there is
  the possibility to compute a solution not worse
  than $\phi$.
  \item[(Valued-)ask] The semantics of the (valued-)ask is extended in
  a way similar to what we have done for the (valued-)tell action. This
  means that to apply the rule we need to check if the store $\sigma$
  entails the constraint $c$ but also if the store is ``consistent
  enough" w.r.t. the threshold ($\phi$ or $a$) set by the programmer.
  \item[Nondeterminism and parallelism] The composition
  operators $+$ and $\|$ are not modified w.r.t. the classical
  one: a parallel agent will succeed if all the agents succeed;
  a nondeterministic rule chooses any agent whose guard succeed.
  Note that at this stage no distinction between the {\em
  don't know} or {\em don't care} observational behaviour is
  made.
  \item[Hidden variables] The semantics of the existential
  quantifier is similar to that of CCP by
  using the notion of {\em freshness} of the new variable added to
  the store.
  \item[Procedure calls] The semantics of the procedure call is not modified
  w.r.t. the classical one. The only
  difference is the different use of the diagonal constraints to
  represent parameter passing.
\end{description}

The transition rules of scc could be applied to both eventual and
atomic tell/ask interpretation. In fact, while the generic tell/ask
rule represents an atomic behaviour, by setting $\phi = 0$ or $a=0$
we obtain their {\em
eventual} version. \\
Notice that, by using an eventual interpretation, the transition
rules of the scc become the same as those of cc (with an eventual
interpretation too). This happens since, in the eventual version,
the tell/ask agent never checks for consistency and so the soft
notion of $\alpha$-consistency does not play any role.


\section{Fairness as a Qos Measure for SOAs} \label{sec:fair}

The fairness attribute we investigate represent the ``\emph{ethical
acceptance of the service}" or the ``\emph{perceived equity}" of the
service. This is clearly a ``\emph{subjective}" quality of the
service and cannot be measured by the service provider just looking
at one of the user, since all the user need to be considered
together. Let us better explain our idea with an example. Suppose
the service we consider is to have access to a table in a
restaurant, and to be served by the waiter. A fair service will pay
attention to the waiting time of the customer, and will try to
provide the service with the ``\emph{same}" waiting time. Indeed, in
order to score the quality of the service, a given user will not
only evaluate only his (or her) own waiting time, but will also
compare this waiting time with the waiting time of the other
customers. This is what we mean for ``perceived" quality of the
service: if a user will wait some more, but will notice that no one
will be served will be less disappointed than seeing other client
sit at a table before him. \\
Fairness in computer science is often related to concurrent systems
and to the guarantee that ``\emph{eventually}" each user will be
served. Our goal is to obtain both perceived equity and fairness as
defined in computer science.

We can represent the users requiring a web service with the (soft)
concurrent constraint framework where the concurrent agents can be
seen as the clients requesting a service to the provider. Each
client has a own measure of QoS that depends by the level of
fairness that the user assigns to the service. The aim of the
provider is instead to offer his service in a fair way, in order to
maximize the QoS level of each customer.

 To model this situation with our CC framework we need a
\textit{fair method of selection} of parallel agents: we propose the
new parallel operator ($\parallel_{m}$), in order to deal with a finite number of agents.\\
We clarify these notions with an example: suppose that more clients
want to obtain a service (offered by a service provider) and this
cannot be provided at the same time to all clients. The service
provider has to select which agent can be satisfied first. In this
case it needs a criterion for selecting one agent rather than
another: this criterion should be as fair as possible to guarantee
to all the customers the same level of perceived quality of the
service. To model this scenario with the CC framework, we extend it
with a new parallel operator, in order to obtain Fair Concurrent
Constraint Programming.

\section{Fair Concurrent Constraint Programming}
In this section we present how to use a quantitative metric to
provide a more accurate way to establish which of the agents
involved in a parallel execution can succeed. The metric we use is
the one proposed by a \emph{Fair carpooling scheduling algorithm}~\cite{carpooling}, which is described more in detail below.

\subsection{The Fair Carpool Scheduling Algorithm}
Carpooling consists in sharing a car among a driver and one or more
passengers to divide the costs of the trip. We want a scheduling
algorithm that will be perceived as fair by all the members as to
encourage their continued participation.\\ Let $U$ a value that
represents the total cost of the trip. It is convenient to take $U$
to be the least common multiple of $[1,2, \ldots, m]$ where $m$ is
the largest number of people who ever ride together at a time in the
carpool. Since in a determined day the participans can be less than
$m$, we define $n$ as the number of participants in the carpooling
in a given day ($n \in
[1 \ldots m]$). \\

Each day we calculate the passengers and driver's scores. This
situation can be represented with a table which cols contains the
members of the carpooling and inside the rows we insert the days;
the intersection contains the score of each participant. \\ The
first row has value zero for each member; in the following days the
driver will increase his score of $U(n-1)/n$ , while the remaining
$n-1$ passengers decrease their score of $U/n$.

 As proved in~\cite{carpooling} this
algorithm is fair; a finite computation is \textit{Carpooling-fair}
if for each enabled agent (or process) $A_{i}$ there is a number $N$
such that at each time $t$, the number of times that $A_{i}$ is
executed ($E$) differs from his \emph{ideal number} of executions
$I$ in absolute value than no more than $N$. The ideal number of
executions represents the number of times each agent wishes to be
executed.
 Formally we have: $ \forall \ A_{i} \ \exists \ N \ \textrm{ s.t. } \ |E-I| \leq N$.
\subsection{The New Parallel Operator} \label{sec:parallelm} In order
to obtain a Fair version of the parallel execution of a finite
number of agents, we modify the parallel operator $\|$ in the CC
semantic by replacing it with the $\|_{m}$ operator, which is able
to deal with a set of $m$ agents. Since with the classical parallel
operator only two concurrent agents are represented in the
Parallelism rule, it is
not possible to express fairness among more than 2 agents at the same level. \\
The syntax of the CC language will be modified by replacing the
$\parallel$ operator with $\parallel_{m} (A_{1}, \ldots ,A_{m})$.
Moreover we develop the new transition rule, which is able to
operate with the metrics of the carpooling algorithm.We introduce an
array $\vec{k}$ and a set of conditions over $\vec{k}$, $<cond(k)>$
for the selection of the agent that we want to execute. The array
$\vec{k} $ (with $m$ elements) allows us to keep track of the
actions performed by each agent during the computational steps (we
use the notation $\langle A_i, \vec{k}, \sigma \rangle$ instead of $
\langle A_{i}, \sigma \rangle $). In this way we can associate a
value $\vec{k}[i]$ to each agent $A_{i}$. The rule considers among
the $m$ agents, the $n$ enabled agents (that we indicate with the
notation: $A_{l_{1}} \rightarrow, \ldots, A_{l_{n}}\rightarrow$).\\
With reference to the carpooling algorithm, we represent the driver
with the agent $A_{l_{i}}$, while the passengers are the remaining
($n-1$) enabled agents. \\
Let $n \leq m$ represents the number of enabled agents and
$A_{l_{i}}$ (with $i=[1,\ldots,n]$) the agent enabled and executed.
$A_{l_{i}}$ increases his score of $U(n-1)/n$, while the $n-1$
enabled and not executed agents decrease their score of $U/n$.
Initially all the $m$ elements of the array $\vec{k}$ are equal to
$0$.\\
We define $\alpha_{n}=U(n-1)/n$ and $\beta_{n}=U/n$. These values
are used in the following way: we add $\alpha_{n}$ to the previous
value ($\vec{k}[l_{i}]$) of the agent $A_{l_{i}}$; we subtract
$\beta_{n}$  to the previous value ($\vec{k}[l_{j}]$) of the other
agents $A_{lj} \quad \forall j \in [1, \ldots ,n], \quad j \neq i $.

In this case we use the condition $<cond(k)>= \vec{k}[l_{i}] \leq
\vec{k}[l_{j}]$ that allows the (enabled) agent with a lower score
to evolve. We update the scores in the variable $\vec{k}$ in the
following way: the agent $A_{l_{i}}$ (that is the agent which
performs the execution) increases his previous score of
$\alpha_{n}$; the ($n-1$) enabled agents decrease their score of
$\beta_{n}$, while the scores of the not enabled agents remain
unchanged. With the parameters provided by the carpooling algorithm
we obtain the Carpooling-fairness rule:


\begin{table}[hbt]
\label{table:carpooling} \caption{Carpooling fairness---parallelism
(1) rule.}
\begin{center}
\begin{tabular}{c}
\hline
\\
$
\inference{ A_{l_{1}} \rightarrow, \ldots, A_{l_{n}} \rightarrow\\
\vec{k}[l_i] \leq \vec{k}[l_j] \quad \forall \quad  j=1,\ldots,n
\quad <A_{li},\sigma> \rightarrow <A_{l_{i}}',\sigma'> }
{<||_{m}(A_{1},\ldots,A_{l_{i}},\ldots,A_{m}), \vec{k},\sigma>
\rightarrow
<||_{m}(A_{1},\ldots,A_{l_{i}}',\ldots,A_{m}),\vec{k}',\sigma'>}
$
\\
\\
$
\vec{k}[x]'  = \left\{
\begin{array}{ll}
\vec{k}[x] & \textrm{ if } x=1,\ldots,m \\
\vec{k}[x] + \alpha_{n} & \textrm{ if }  x=l_{i} \\
\vec{k}[x] - \beta_{n}  & \textrm{ if } x=l_{j}  \qquad j \neq i
\end{array}\right.
$
\\
\\
\hline
\end{tabular}
\end{center}
\end{table}

At each computation we select an agent according its value in the
array $\vec{k}$ and then we update it in a way that at the following
step, another agent has to be executed. This comply with the
carpooling-fairness we defined.


We use the same criterion also for the Parallelism (2) rule:
\begin{table}[hbt]\label{table:carpooling2}
\caption{Carpooling fairness---parallelism (2) rule.} 
\begin{center}
\begin{tabular}{c}
\hline
\\
$
\inference{ A_{l_{1}} \rightarrow, \ldots, A_{l_{n}} \rightarrow
\\
\vec{k}[l_i] \leq \vec{k}[l_j] \quad \forall \quad  j=1,\ldots,n
\quad <A_{li},\sigma> \rightarrow <success,\sigma'> }
{<||_{m}(A_{1},\ldots,A_{l_{i}},\ldots,A_{m}), \vec{k},\sigma>
\rightarrow
<||_{m-1}(A_{1},\ldots,A_{l_{i}}-1,A_{l_{i}}+1,\ldots,A_{m}),\vec{k}',\sigma'>}
$
\\
\\
$
\vec{k}[x]'  = \vec{k}[x] \quad \backslash \quad \vec{k}[l_{i}]
$
\\
\\
\hline
\end{tabular}
\end{center}
\end{table}

If agent $A_{l_{i}}$ succeed, is deleted from the rule together with
the corresponding element $\vec{k}[l_{i}]$, while the other values
within the array $\vec{k}$ remain unchanged. Consequently the value
$m$ of the operator $\parallel_{m}$ is continuously decreased of 1
every time an agent performs a \emph{success}. We obtain therefore a
final computation with a single agent and a single element on the
array that reinstate the original rule: $ \langle
\parallel_{1}(A_{1}),\vec{k}[1], \sigma \rangle \to \langle A_{1},
\sigma \rangle $.
\\

The new definition of the $\parallel_{m}$ operator permit us to
maintain fairness step by step; in fact the provider of the service
selects the agent according a well-defined fair criterion.  Notice
that as it is based on the classical CC framework, this actual
extension of the parallel operator does not consider the preference
level/fuzzyness of the
constraint inserted into the store with a tell action. \\
In fact, to associate a level of preference to the constraints we
need to use a semiring-based structure (see next section).

\section{Using Fair (S)CC for SLA Negotiation}
 We use our fair extension of (S)CCP language to model a formal agreement. A
Service Level Agreement~\cite{sla} is an agreement regarding the
guarantees of a web service. The parties involved in the SLA are the
Service provider and the customers. The goal is to accomplish the
requests of all the agents by satisfying their QoS requirements in a
fair way, that is, we want that each customer is equally satisfied
with respect to the other customers requesting the same service.

Suppose there are three customers (agents $A_{1}, A_{2}, A_{3}$)
that
want to request a service.\\
We model this scenario with the SCC language (therefore we use soft
constraints) where each constraint has an associated preference
level. We represent the requests of the service with the parallel
execution of the three agents performing tell actions (as showed
below) to the store:
\\
\begin{center}
\hrule \vskip2pt
\begin{eqnarray*}
& A_{1} & \equiv
tell(c_{1}) \rightarrow^a tell(c_{2}) \rightarrow^a
tell(c_{3})\rightarrow^a success\\
& \parallel & \\
& A_{2} & \equiv tell(c_{4}) \rightarrow^a tell(c_{5})\rightarrow^a success\\
& \parallel & \\
& A_{3} & \equiv tell(c_{6}) \rightarrow^a tell(c_{7})\rightarrow^a
tell(c_{8}) \rightarrow^a tell(c_{9}) \rightarrow^a success
\end{eqnarray*} \vskip2pt
\hrule
\end{center}
\medskip

In order to guarantee equity when is not possible for the service
provider to offer the service contemporaneously, we have to select
the agents according their level of preference; this is associated
to each constraint and represents the subjective level of fairness
requested by each agent.
\\

We can use the new $\parallel_{m}$ operator to deal with soft
constraints instead of crisp ones; we consider Soft Concurrent
Constraint that permits to operate with preferences associated to
the constraints. In this case, we want to select the agent with a
lower level of preference; in the soft case the array $\vec{k}$
previously defined in the general rule contains soft constraints
(unlike values used in the crisp case) added by the Agents which
perform a tell action. This array is divided into sections: for each
agent $A_{i}$ there is a section $\vec{k}[i]$ that contains the
agent's constraints. These are combined through the $\otimes$
operator. For example, if Agent i performs the action: $tell(c)$, we
will modify the section $\vec{k}[i]$, obtaining $\vec{k}=[k_{1},
\ldots, k_{l_{i}} \otimes c, \ldots, k_{m}]$. Notice that in the
soft case we don't use counters or values (as $\alpha$ and $\beta$
defined in the carpooling algorithm),
but we just perform the combination of the constraints of the agent that succeed. \\
Below we show the modified rule of the parallelism operator that
uses the array of soft constraints, and as condition
$Con=\vec{k}[l_{i}] \sqsubseteq \vec{k}[l_{j}]$ :

\begin{table}[hbt]
\begin{center}
\begin{tabular}{c}
\textrm{$\|_{m}$ - parallelism (1) with soft constraints.}\\
\hline
\\
$
\inference{ A_{l_{1}} \rightarrow, \ldots, A_{l_{n}} \rightarrow
\\
\vec{k}[i] \sqsubseteq \vec{k}[j]  \qquad <A_{li},\sigma>
\rightarrow <A_{l_{i}}',\sigma'> }
{<||_{m}(A_{1},\ldots,A_{l_{i}},\ldots,A_{m}), \vec{k},\sigma>
\rightarrow
<||_{m}(A_{1},\ldots,A_{l_{i}}',\ldots,A_{m}),\vec{k}',\sigma'>}
$
\\
\\
$
\vec{k}[x]' \left \{
\begin{array}{ll}
\vec{k}[x]  &  \textrm{ if }  x=1,\ldots,m \\
\vec{k}[x] \otimes c & \textrm{ if }  x=l_{i}  \\
\vec{k}[x]  & \textrm{ if } x=l_{j}  \quad j \neq i
\end{array}\right.
$
\\
\\
\hline
\\
\end{tabular}
\end{center}
\end{table}

This rule means that when the agent with the lower level of
preference ($A_{l_{i}}$) is selected (since the guard
$\vec{k}[l_{i}] \sqsubseteq \vec{k}[l_{j}]$ permits it), it combines
its constraint ($c$) with the element $\vec{k}[l_{i}]$ obtaining
$\vec{k}[l_{i}] \otimes c$; therefore its level of preference, that
is his QoS measure, increases (this means that only the agent
selected changes his preference level) and this constraint is
inserted in the store $\sigma$; in this way we obtain a new store
$\sigma^{'}$ that
corresponds to the old store combined with the new constraint: $\sigma^{'}=\sigma \otimes c$. \\
 Notice that if the computation $A_{l_{i}} \to A_{l_{i}}'$ does not change the store,
 the computation evolves in a normal way, without affecting the vector $\vec{k}$.
In the next step another agent is selected and its constraint
combined with the previous constraints of the array $\vec{k}$, then
it is added to the store and so on for the successive steps. \\

With the new transition rule, the provider will select the agent
 with the lower preference level, then it will combine its
constraint through the $\otimes$ operator with the previous
constraint in $\vec{k}$ and then to the store; in the example above,
if at first step the enabled agent (with the lower soft constraint)
is $A_{1}$, it will insert its constraint to the store, in the
section $\sigma_{1}$, obtaining $\sigma_{1} \otimes c_{1}$; after
this it can succeed to the following step (tell($c_{2}$)). In the
next step another agent is selected and its constraint added to the
store and so on for the successive steps.
\\
This rule is fair according the notions defined in section
\ref{sec:fair}, because considers both the perceived equity, that
is, each agent is served according a comparison with the preferences
of the others, and fairness as the eventual execution of the agents;
in fact, each enabled agent will be necessarily executed.\\ As each
agent considers the service fair, this parameter will contribute to
the increasing of the Quality of Service measure.

\section{Related Work}
In this paper we modeled web services with the soft concurrent
constraint language, considering fairness as a QoS parameter. In~\cite{rw1}, where is studied air channel sharing in WLAN access
networks, the authors propose fairness and quality of service as
separeted issues in allocating wireless channels. In~\cite{rw2}, is
studied how a fair access in IEEE 802.11 bear its impact on quality
of service. In~\cite{rw3} A Web services implementation is
presented; the paper introduces a flexible framework to support fair
non-repudiable B2B interactions based on a trusted delivery agent.

\section{Future Work}
In this article we used soft constraints as a fairness measure for
QoS and we modeled the Service Level Agreement with the Soft
Concurrent Constraint language. As future work we are going to apply
the proposed formal technique to running web service example
monitored for quality of service. In this way we can establish how
the present approach on including fairness as a criteria actually
influences the selection of a service by a service requester.

In economics the concept of fairness (or equity) is related instead
to a function that represent the utility of each user. The concept
of equity is in this case represented by some indexes that give a
quantitative measure of fairness. There are various indexes for
measuring economic inequality. The Gini coefficient is a measure of
statistical dispersion most prominently used as a measure of
inequality of income distribution or inequality of wealth
distribution. It is defined as a ratio with values between 0 and 1:
a low Gini coefficient indicates more equal income or wealth
distribution, while a high Gini coefficient indicates more unequal
distribution. We plan to focus on fairness by paying attention to
these economic aspects and to the use of social welfare functions
and indexes.
\\

\bibliographystyle{eptcs} 

\end{document}